\documentclass[aps,prc,twocolumn,superscriptaddress]{revtex4}

\usepackage{graphicx}

\begin{document}

\title{Study of $^{124}$Sn+$^{136}$Xe fusion-evaporation: analysis of a rare-event experiment}

\author{B. Avez}
\affiliation{CEA, Centre de Saclay, IRFU/Service de Physique Nucl\'eaire, F-91191 Gif-sur-Yvette, France }

\author{A. Drouart}
\affiliation{CEA, Centre de Saclay, IRFU/Service de Physique Nucl\'eaire, F-91191 Gif-sur-Yvette, France }

\author{Ch. Stodel}
\affiliation{GANIL, Bd Henri Becquerel, BP 55027, F-14076 Caen Cedex 5, France}

\author{C. Simenel}
\affiliation{CEA, Centre de Saclay, IRFU/Service de Physique Nucl\'eaire, F-91191 Gif-sur-Yvette, France }

\author{J. Alcantara}
\affiliation{GANIL, Bd Henri Becquerel, BP 55027, F-14076 Caen Cedex 5, France}

\author{M. Adamczyk}
\affiliation{Smoluchowski Institute of Physics, Jagiellonian University, Krakow, Poland} 

\author{P. Banka}
\affiliation{Smoluchowski Institute of Physics, Jagiellonian University, Krakow, Poland}

\author{E. Bonnet}
\affiliation{GANIL, Bd Henri Becquerel, BP 55027, F-14076 Caen Cedex 5, France}

\author{E. Cl\'ement}
\affiliation{GANIL, Bd Henri Becquerel, BP 55027, F-14076 Caen Cedex 5, France}

\author{R. Dayras}
\affiliation{CEA, Centre de Saclay, IRFU/Service de Physique Nucl\'eaire, F-91191 Gif-sur-Yvette, France }

\author{C. Force}
\affiliation{GANIL, Bd Henri Becquerel, BP 55027, F-14076 Caen Cedex 5, France}

\author{C. Golabek}
\affiliation{GANIL, Bd Henri Becquerel, BP 55027, F-14076 Caen Cedex 5, France}

\author{A. Gonciarz}
\affiliation{Smoluchowski Institute of Physics, Jagiellonian University, Krakow, Poland}

\author{S. Grevy}
\affiliation{GANIL, Bd Henri Becquerel, BP 55027, F-14076 Caen Cedex 5, France}

\author{K. Hauschild}
\affiliation{CSNSM, CNRS/IN2P3, F-91405 Orsay Campus, France}

\author{D. Jacquet}
\affiliation{IPNO, CNRS/IN2P3, Universit\'e Paris-Sud 11, F-91406 Orsay Cedex, France}

\author{A. Korichi}
\affiliation{CSNSM, CNRS/IN2P3, F-91405 Orsay Campus, France}

\author{T. Kozik}
\affiliation{Smoluchowski Institute of Physics, Jagiellonian University, Krakow, Poland}

\author{P. Lazko}
\affiliation{Smoluchowski Institute of Physics, Jagiellonian University, Krakow, Poland}

\author{M. Morjean}
\affiliation{GANIL, Bd Henri Becquerel, BP 55027, F-14076 Caen Cedex 5, France}

\author{A. Popeko}
\affiliation{Flerov Laboratory of nuclear reactions, JINR, Joliot-Curie str.  6, Dubna, 141980, Russia}

\author{T. Roger}
\affiliation{GANIL, Bd Henri Becquerel, BP 55027, F-14076 Caen Cedex 5, France}

\author{M.-G. Saint-Laurent}
\affiliation{GANIL, Bd Henri Becquerel, BP 55027, F-14076 Caen Cedex 5, France}

\author{Z. Sosin}
\affiliation{Smoluchowski Institute of Physics, Jagiellonian University, Krakow, Poland}

\author{B. Sulignano}
\affiliation{CEA, Centre de Saclay, IRFU/Service de Physique Nucl\'eaire, F-91191 Gif-sur-Yvette, France }

\author{Ch. Theisen}
\affiliation{CEA, Centre de Saclay, IRFU/Service de Physique Nucl\'eaire, F-91191 Gif-sur-Yvette, France }

\author{A. Wieloch}
\affiliation{Smoluchowski Institute of Physics, Jagiellonian University, Krakow, Poland}

\author{A. Yeremin}
\affiliation{Flerov Laboratory of nuclear reactions, JINR, Joliot-Curie str.  6, Dubna, 141980, Russia}

\author{B. Yilmaz}
\affiliation{Physics Department, Ankara University, 06100 Tandogan Ankara,  T\"urkiye}

\author{M. Zielinska}
\affiliation{Heavy Ion Laboratory, Warsaw University, Warsaw, PL-02097, Poland}

\begin{abstract}
Fusion-evaporation in the $^{124}$Sn+$^{136}$Xe system is studied using a high intensity xenon beam provided by the Ganil accelerator and the LISE3 wien filter for the selection of the products. 
Due to the mass symmetry of the entrance system, the rejection of the beam by the spectrometer was of the order of $5\times10^8$.
We have thus performed a detailed statistical analysis to estimate random events and to infer the fusion-evaporation cross sections. No signicant decay events were detected and upper limit cross sections of 172~pb, 87~pb and 235~pb were deduced for the synthesis of $^{257}$Rf, $^{258}$Rf and $^{259}$Rf, respectively.
\end{abstract}

\maketitle

\section{Introduction}
In most experiments, superheavy elements are formed via mass asymmetric fusion-evaporation reaction, with a light projectile impinging on a heavy target ~\cite{hof00,oga06,hof07,mor07}. Symmetric entrance sytems (where projectile and target have similar masses) are known to exhibit small fusion cross-sections ~\cite{mor91,gag84,faz05}. Nevertheless, such systems may have interesting properties, specially when aiming at very heavy elements. 
First, such systems can give access to specific isotopes. This is the case for the $^{136}$Xe+$^{136}$Xe that leads to the $^{272}$Hs compound nucleus which has 6 more neutrons than the one reached through the $^{58}$Fe+$^{208}$Pb. Moreover, the excitation energy is low ($E^*=-5$~MeV at the Bass barrier, so we can reach the 1 or 2 neutrons evaporation channels only), leading to a rather neutron rich evaporation residue. 
In addition, future ISOL radioactive beam facilities, like SPIRAL2 \cite{lew08} will provide high intensity neutron-rich beams from uranium fission. In SPIRAL2, the beams provided with the highest intensities will be neutron-rich Kr and Xe isotopes. The latter may be interesting for the production of heavy and superheavy elements in mass quasi-symmetric fusion reactions. In particular, the use of neutron-rich beams in such reactions could form new isotopes closer to the $\beta$-stability line than with stable beams. \\
Some experiments have been dedicated to the synthesis of very heavy elements through symmetric reactions. The $^{136}$Xe+$^{136}$Xe system was studied in Ref.~\cite{oga07}, and an upper limit cross-section of 4~pb was obtained. In the present work, we investigate the formation of fusion-evaporation residues in the $^{124}$Sn+$^{136}$Xe system. 
The choice of this reaction is motivated by the relatively light (and then less fissile) compound nucleus $^{260}$Rf and the fact that two closed shells (Z=50 and N=82) are expected to favor fusion by limiting quasi-fission in the entrance channel~\cite{faz05,sim11}.
From a technical point of view, it is relatively easy to build solid tin targets. 
Previous investigations of this system led to respective fusion-evaporation cross-section upper limits of 200~pb at an excitation energy $E^*=26$~MeV~\cite{pop01} and 1~nb at $E^*=15$~MeV~\cite{ike02}. For comparison, in the reaction $^{50}$Ti+$^{208}$Pb leading to the $^{258}$Rf compound nucleus, the cross sections of the 1n and 2n evaporation channels have been measured to be 10 and 12~nb respectively~\cite{hes97}.\\
From an experimental point of view, 
symmetric reactions have the advantage of a forward focused kinematics compared to asymmetric systems in normal kinematics. The transmission of the residues through a separator device is enhanced and they are produced with higher kinetic energy; thus their detection in an implantation detector is facilitated. 
The main drawback of symmetric reaction is the relatively close velocities of the residues and of the beam ions, which make them harder to separate. For this reason, symmetric reactions usually lead to the detection of random events producing a relatively high background in the spectra as compared to asymmetric systems in normal kinematics. Here, we present a practical statistical analysis that takes very rigorously into account the possibility of random correlations. This kind of analysis is very adapted to any super heavy element production reaction where such conditions of very low signal to noise ratio are encountered.

\section{\label{sec:exp_details} The Fulis experimental setup} 

\begin{figure*}
\includegraphics[angle=0,width=8cm]{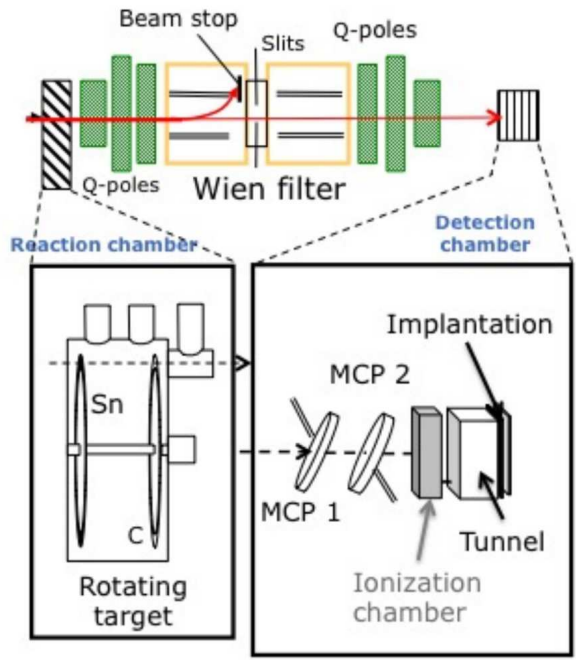}
\caption{Experimental setup.}
\label{fig:setup}
\end{figure*}

The experiment took place at GANIL, at the end of the LISE 3 beam line~\cite{ann92}. 
The figure~\ref{fig:setup} gives a scheme of the set-up. 

\subsection{Projectiles and targets}

The CSS1 cyclotron of GANIL provides a 4.61~MeV/u $^{136}$Xe beam with an intensity of $6.9\times10^{11}$~pps. 
The total dose on target is $1.07\times10^{17}$ nuclei. 
The isotopically enriched $^{124}$Sn targets are 400~$\mu$g/cm$^2$ thick, 
with a 40~$\mu$g/cm$^2$ carbon backing on the entrance side and 5~$\mu$g/cm$^2$ on the exit side. 
Center-of-mass (c.m.) energy at mid target is $E_{c.m.}=293.9$~MeV, for a compound nucleus excitation energy around $E^*\approx20$~MeV. 
The targets are mounted on 18 separated sectors of a 36~cm diameter rotating wheel. 
16~cm downstream, a similar wheel hosts 40~$\mu$g/cm$^2$ carbon strippers to equilibrate the evaporation residue charge states. 
The wheels turned at 2000~rpm during most of the experiment. 

\subsection{Transmission in the Wien filter}

The compound nucleus (CN) is $^{260}$Rf and standard statistical calculations predict an evaporation from one to three neutrons.
The ion-optical line is simulated with the Zgoubi ray-tracing code~\cite{meo99}.
These simulations have been validated in previous experiments \cite{sto07,cha07}. 
Following the targets and strippers, a triplet of quadrupoles focuses the nuclei at a focal point at the middle of the Wien Filter (WF). 
This WF consists in two identical symmetric parts. They both combine a uniform magnetic field with an orthogonal electric field. 
If the fields have a ratio $E/B=v_{ref}$, the ions that have a velocity equal to $v_{ref}$ are not deviated, whatever their charge state. 
Ions with other velocities are deviated. 
In a standard mode, $v_{ref}$ is chosen to be equal to the velocity of the evaporation residues, which is 1.53~cm/ns in our case. 
The beam ions are faster (2.93~cm/ns) and, thus, are deviated toward a copper plate placed at the middle of the WF. 
This plate is water-cooled to sustain the heat deposit. 

A direct beam with the same velocity as the residues $v_{ref}=1.53$~cm/ns after stripping from the carbon foils was used to check the optical tuning. 
The charge state distribution was estimated using the model of Ref.~\cite{bar93}
The optimum transmission is calculated to be $90\%$ for the central charge state $\bar{q}=43e$ 
with an electric field of $E=200$~kV/m and a magnetic field $B=150$~Gauss in the WF. 
With this configuration, the rejection of the beam was measured to be $10^7$ which is too small to implement an efficient recoil decay tagging technique. 
The rejection was improved up to $5\times10^8$ by increasing the WF magnetic field to $B=190$~Gauss, which is associated to a lower $v_{ref}$.
That led to a counting rate of 1300~counts/sec on the full implantation detector. 
As there is no free lunch, the CN transmission was also reduced to $45\%$ for the central charge state, resulting to a total transmission of $38\%$ for all charge states. 
After the WF, another triplet of quadrupoles is set in order to refocus the transmitted nuclei on the final implantation detector.

\subsection{Detection system}

The detection system is first composed of two emissive foils detectors. 
These detectors are made of aluminized Mylar foils of 0.9~$\mu$m thickness. 
When they cross the foil, the heavy ions emit secondary electrons that are accelerated with an electrostatic field toward micro-channel plates (MCP). 
Their fast signal gives a time reference for the measurement of the time-of-flight (TOF). 
In addition, a signal in one of the MCP is used as a veto for the detection of alpha or fission decay in the implantation detector. 
The efficiency of this veto is above $99.9\%$. 

The BEST (Box Electron Spectroscopy after Tagging) detector closely follows the two emissive foils. 
BEST includes a double-sided silicon-stripped detector (DSSSD). 48 strips on each side (vertical and horizontal) give the (x,y) position. 
The detection surface of the DSSSD is $50\times50$~mm$^2$, with a thickness of 300~$\mu$m. 
Its resolution is 60~keV (FWHM) around 5.5~MeV, measured with a 3-$\alpha$ source ($^{239}$Pu, $^{241}$Am, $^{244}$Cm). 
Relatively to this calibration, the energy of the alpha decays occurring within the detector is corrected from the known dead zone~\cite{cha06}. 
For each implanted ion, the TOF and energy are recorded for a primary selection of the evaporation residues: 
for a similar velocity, the kinetic energy of the beam is twice as small as the one of the residue. 

Four ''tunnel'' detectors frame the entrance face of the implantation, with their normal perpendicular to the beam axis. 
Each tunnel detector has a surface of $50\times50$~mm$^2$, a thickness of 1~mm of silicon, and contains four independent pixels. 
They detect backward emitted alpha particles or fission fragments emitted from the implantation DSSSD.
This increases the alpha decay detection efficiency from $55\%$ to $80\%$. 
The implantation of a residue is detected at a known $\{x,y\}$ position. 
A subsequent decay (alpha or fission) is characterized by a signal recorded at the same position, and possibly in one of the tunnel detectors. 
To make sure that the associated signals come from subsequent decays and not from implantation, 
the absence of any signal in the two MCP detectors is required. 

During the experiment, we have also tested a scintillation-ionization chamber that was placed after the MCPs, 
but it is not used in the present analysis due to technical problems. The details on this detector can be found in~\cite{wie10}.
The acquisition of an event was triggered either by a signal in the implantation detector (most of the events) or in the tunnel. Finally, the counting rate of around 1~kHz generated an electronic dead-time of about $20\%$. 
It is taken into account in the analysis. Beam intensity is also recorded to calculate the absolute cross-sections.

\section{data analysis}

\subsection{Identification matrix}

\begin{figure}
\includegraphics[angle=0,width=8cm]{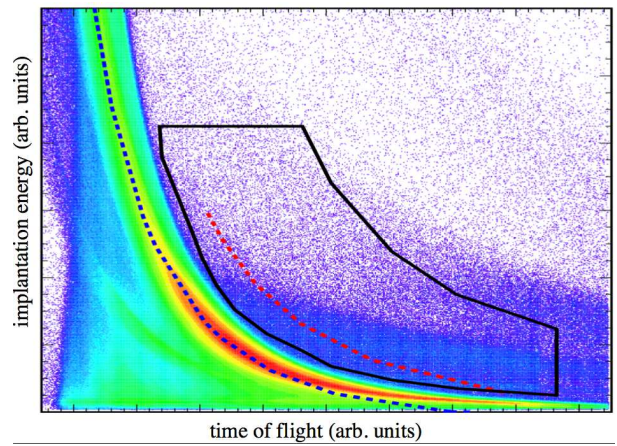}
\caption{Identification matrix of the detected ions.
Implantation energy in the DSSSD as a function of the time of flight between the first emissive foil detector and the implantation. LISE++ calculations of scattered $^{136}$Xe events (blue) and of the residues (red) are shown with dashed lines. The black solid line delimitates the area where fusion-evaporation residues are searched for.}
\label{fig:DE_TOF}
\end{figure}

To identify fusion-evaporation residues, a correlation between the energy of the ions and their TOF is used. 
The energy is given by the DSSSD in coincidence with a signal in the first emissive foil detector. 
The long path between these two detectors also ensures a good resolution for the TOF measurement (close to 1~ns). 
Figure~\ref{fig:DE_TOF} shows the identification matrix obtained from these two quantities. 
Calculations based on the code LISE++~\cite{tar02}, taking into account the energy loss in the different elements of the experimental setup, give the expected position of the transmitted beam-like particles (blue dashed line) and of $^{258}$Rf residues (red dashed line). 
It is clear that the identification matrix is not sufficient to identify fusion-evaporation residues due to the important amount of transmitted beam-like particles. 
In what follows, a very conservative condition (inside the solid line contour) is applied to look for fusion-evaporation residues. 

\subsection{Genetic correlation method}

\begin{table*}
\caption{\label{tab:ER} Decay properties of expected ER (rutherfordium) and their daughter (nobelium) nuclei.}
\begin{tabular}{ccccccc}
\hline\noalign{\smallskip}
nucleus & $t_{1/2}$ & $\alpha$-BR & sf-BR & EC-BR & $E_\alpha$ (keV) & Reference\\
\noalign{\smallskip}\hline\noalign{\smallskip}
$^{257}$Rf & $4.5\pm1$ s & $>89\%$ & $\le3.5\%$ & $11\pm1\%$ & $\{8283\pm18$ ... $9021\pm18\}$ & \cite{hes97} \\
$^{258}$Rf & $14.7^{+1.2}_{-1.0}$ ms & $31\pm11\%$ & $69\pm11\%$ & & $9050\pm30$ & \cite{gat08} \\
$^{259}$Rf & $2.5^{+0.4}_{-0.3}$ s  & $85\pm4\%$ & & $15\pm4\%$ & $8770\pm15,$ $8865\pm15$& \cite{gat08}  \\
$^{253}$No & $1.57^{+0.18}_{-0.15}$ min & $\approx80\%$ & & $\approx20\%$ & $\{8011\pm21$ ...  $8114\pm18\}$ & \cite{hes97,qia09} \\
$^{254}$No & $55\pm$3 s & $90\pm4\%$ & $0.17\pm0.05\%$ & $10\pm4\%$ & $8093\pm14$ & \cite{chu99} \\
$^{255}$No & $3.1\pm$0.2 min & $61.4\pm2.5\%$ & & $38.6\pm2.5\%$ & $\{7620\pm10$ ... $8312\pm9\}$ & \cite{chu99} \\
\noalign{\smallskip}\hline
\end{tabular}
\vspace*{0cm}  
\end{table*}

The next step in searching for fusion-evaporation residues (ER) is to study the radioactive $\alpha$-decay and/or spontaneous fission (sf) after the implantation of an ion in the DSSSD. 
Let us define and label three different types of correlations between the implantation of the ER and its subsequent decay: 
\begin{itemize}
\item $\{ER-\alpha_1\}$ the measurement of at least one signal in the same pixel (or one of its closest neighbours) of the DSSSD, which could be associated to an $\alpha$-decay of a specific ER after an implantation,
\item $\{ER-\alpha_1-\alpha_2\}$ the same with two alphas, i.e., with the subsequent decay of the daughter,
\item and $\{ER-sf\}$ when a signal following the implantation could be associated to a spontaneous fission  event. (Fission fragments deposit much more energy than $\alpha$-decays.)
\end{itemize}
The $\alpha$-decay energies, half-life $t_{1/2}$, and branching-ratios (BR) of every expected ER and its daughter nucleus are given in table~\ref{tab:ER}.

The $\alpha$ and sf-decays are searched for during a time 3$t_{1/2}$ after the implantation or the previous decay.
According the exponential decay law, the probability to have a decay during this time is $95\%$.
A condition for an $\{ER-\alpha_1\}$ event is that the energy of the ''$\alpha$-signal'' is close to known $E_\alpha$ of the ER. 
A window of $E_\alpha\pm200$~keV is used. The same procedure is applied for the decay of the daughter for $\{ER-\alpha_1-\alpha_2\}$ events. 
Fission fragments are searched for in $^{258}$Rf decay only.
Indeed, the sf-BR is negligible in the other nuclei.  
 The lower limit set for the detection of fission fragments in terms of deposited energy is $E_{sf}^{min}=25$~MeV. 

\subsection{Backgroung from random events \label{sec:back}}

\begin{table}
\caption{\label{tab:events} Number of events which could be associated to the implantation of a fusion-evaporation residue, followed by its decay (labelled as ''{\it correlated}''). An estimate of the number of {\it uncorrelated} events is also given (see text).}
\begin{tabular}{ccccc}
\hline\noalign{\smallskip}
&ER &  $^{257}$Rf & $^{258}$Rf & $^{259}$Rf \\
\noalign{\smallskip}\hline\noalign{\smallskip}
&$\{ER-\alpha_1\}$ & 105 & 0 & 22 \\
{\it correlated }&$\{ER-\alpha_1-\alpha_2\}$ & 0 & 0 & 0 \\
&$\{ER-sf\}$  & & 9 & \\
\hline
&$\{ER-\alpha_1\}$ & 103 & 0 & 14 \\
{\it uncorrelated }&$\{ER-\alpha_1-\alpha_2\}$ & 2 & 0 & 0 \\
&$\{ER-sf\}$  & & 12 & \\
\noalign{\smallskip}\hline
\end{tabular}
\vspace*{0cm}  
\end{table}

The rejection of beam-like ions by the WF and the off-line analysis (see Fig.~\ref{fig:DE_TOF}) is not sufficient to ensure that the results of the genetic correlation method is free from background due to random events. 
To estimate such background, the same method is applied to time-uncorrelated events. 
In practice, such events are obtained by looking for  correlations backward in time in the data.
In this case, any recorded correlation is random. 

Table~\ref{tab:events} gives a summary of the number of events associated to the three sets of decays described above. 
None of these events are accompanied by a signal in the tunnel part of the BEST detector. 
The estimate of uncorrelated events by looking for correlation backward in time is also given in table~\ref{tab:events}.
Globally, compatible statistics are obtained for both correlated and uncorrelated events, showing that unambiguous detection of fusion-evaporation events have not been achieved in the present experiment.

\section{Upper limit of the fusion-evaporation cross-section}

The previous analysis leads us to the evaluation of an upper limit for the fusion-evaporation cross-sections. 
The statistical analysis method is first described. Then, it is applied to the present experimental data to determine an upper limit of the number of physical correlations. Finally, these results are used to estimate the upper limits of the fusion-evaporation cross-sections. 

\subsection{Statistical analysis method}

To determine an upper limit for the production rates in the presence of random events, 
we use a statistical analysis method where we assume that both physical (correlated) and random (uncorrelated) events follow Poisson distributions of the type
\begin{equation}
p(N|\lambda) = e^{-\lambda}\,\frac{\lambda^N}{N!}\, ,
\end{equation}
where $\lambda$ is the expected number of events and $p(N|\lambda)$ is the probability to have $N$ events if $\lambda$ is known.
For simplicity, our choice for the unit of time is the total acquisition time during the experiment. 
With this choice, $\lambda$ is equivalent to a (constant) production rate.
We note $\lambda_p$ and $\lambda_r$ the production rate associated to the physical and random events, respectively.

We first present the method in the case where $\lambda_{r}$ is known. Then, we extend the approach to the case where $\lambda_{r}$ is unknown.

\subsubsection{Case where $\lambda_r$ is known}

We follow the approach given by Br\"{u}chle in Ref.~\cite{bru03}. 
During the experiment, we measure a total number of events $K$, which is a sum of $N$ physical events and $R$ random events: $K=N+R$. 
Knowing the probability law governing the statistics of random events, we can evaluate the probability of having $R$ random events (i.e., the probability to measure $R$ events, knowing $K$ and $\lambda_r$):
\begin{eqnarray}
p(R|K,\lambda_r) &=& \frac{e^{-\lambda_r}\,\frac{\lambda_r^R}{R!}}{\sum_{n=0}^{K}e^{-\lambda_r}\, \frac{\lambda_r^n}{n!}} \nonumber \\
&=& \frac{\lambda_r^R}{R!\,\sum_{n=0}^{K} \frac{\lambda_r^n}{n!}}.
\label{eq:p(R|K,L_r)}
\end{eqnarray}
Of course, because $N=K-R$, we have $p(N|K)=p(R|K)$.

The density of probability $\mathcal{P}(\lambda_p|N)$ to have a distribution of physical events with a production rate $\lambda_p$ when $N$ physical events have been obtained is given by the same Poisson distribution 
\begin{equation}
\mathcal{P}(\lambda_p|N)=p(N|\lambda_p)=e^{-\lambda_p}\, \frac{\lambda_p^N}{N!}.
\label{eq:mP(L_p|P)}
\end{equation}

In fact, we don't know $N$ but only its probability distribution $p(N|K,\lambda_r)$.
Then, Eqs.~(\ref{eq:p(R|K,L_r)}) and~(\ref{eq:mP(L_p|P)}) give 
\begin{eqnarray}
\mathcal{P}(\lambda_p|K,\lambda_r) 
&=& \sum_{N=0}^Kp(N|K,\lambda_r)\, \mathcal{P}(\lambda_p|N) \nonumber \\
&=& \sum_{N=0}^K\frac{\lambda_r^{K-N}}{(K-N)!\, \sum_{n=0}^{K}\frac{\lambda_r^n}{n!}}\,e^{-\lambda_p}\, \frac{\lambda_p^N}{N!}.
\label{eq:L_rknown}
\end{eqnarray}

\subsubsection{Case where $\lambda_r$ is unknown}

Eq.~(\ref{eq:L_rknown}) would be sufficient to determine an upper limit of the production rate of physical events if the statistical distribution of random events (i.e., $\lambda_r$) was known. 
However, only one measurement of the latter is made by searching for events backward in time (see section~\ref{sec:back}). 
Let the result of this measurement be $R_b$.
The density of probability for the rate of random events to be $\lambda_r$ is 
\begin{equation}
\mathcal{P}(\lambda_r|R_b) = e^{-\lambda_r}\, \frac{\lambda_r^{R_b}}{R_{b}!}.
\end{equation}

The probability to find $N$ physical events is then 
\begin{eqnarray}
p(N|K,R_{b}) &=&  \int_0^\infty d\lambda_r  \mathcal{P}(\lambda_r|R_b)\, p(N|K,\lambda_r) \nonumber \\
&=& \int_0^\infty d\lambda_r
 \frac{{e^{-\lambda_r}\, \lambda_r^{K-N+R_b}}}{{R_b! (K-N)!}\,\sum_{n=0}^K \frac{\lambda_r^n}{n!}}.\nonumber \\
 \label{eq:p(P)}
\end{eqnarray}
Similarly to Eq.~(\ref{eq:L_rknown}), the density of probability to have a production rate $\lambda_r$ of physical events is
\begin{equation}
\mathcal{P}(\lambda_p|K,R_b) 
= \sum_{N=0}^K  p(N|K,R_b) \, e^{-\lambda_p}\, \frac{\lambda_p^N}{N!}.
\label{eq:L_runknown}
\end{equation}

\subsection{Application to the present experimental data}

\subsubsection{Physical and random events probability distributions}

\begin{figure*}
\includegraphics[angle=0,width=12cm]{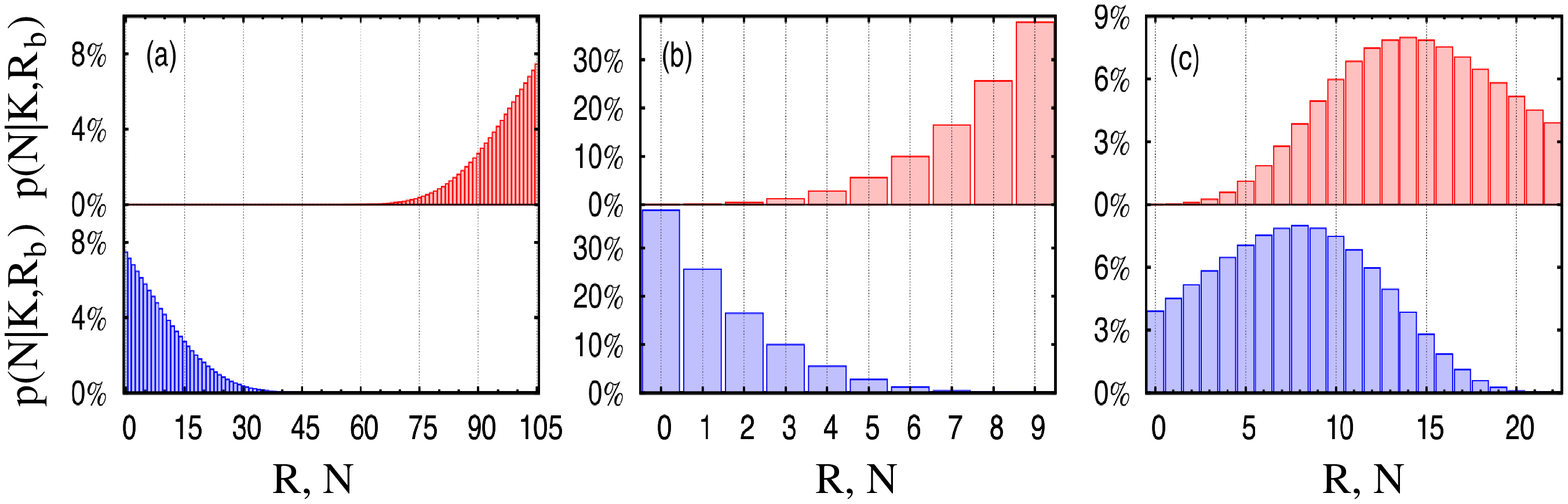}
\caption{Probability distributions for the measurement of  $R$ random events (upper panels) and $N$ physical events (lower panels) for (a)  $\{^{257}$Rf$-\alpha_1\}$, (b) $\{^{258}$Rf$-sf\}$, and (c) $\{^{259}$Rf$-\alpha_1\}$ correlations.}
\label{fig:p(P)}
\end{figure*}

The probability distributions of physical and random events, knowing the total number of events $K=N+R$ and one  estimate of the random events $R_b$, is given by Eq.~(\ref{eq:p(P)}). Using the numbers of table~\ref{tab:events}, these distributions are plotted in Fig.~\ref{fig:p(P)} for the  $\{^{257}$Rf$-\alpha_1\}$ (left), $\{^{258}$Rf$-sf\}$ (middle), and $\{^{259}$Rf$-\alpha_1\}$ (right) correlations, which are the only cases with $K\ne0$. The highest probability is for $N=0$ for the $\{^{257}$Rf$-\alpha_1\}$ and $\{^{258}$Rf$-sf\}$ correlations. For $\{^{259}$Rf$-\alpha_1\}$, however, the highest probability corresponds to the measurement of $N=8$ physical events. This is due to the difference between the number of correlations $K=22$ and the estimated number of random events $R_b=14$ (see table~\ref{tab:events}). However, the probability associated to $N=0$ is only half the maximum one [see Fig.~\ref{fig:p(P)}(c)], and a claim for the observation of $^{259}$Rf residues in this experiment would not be physically grounded. This is all the more so true as no $^{259}$Rf-$\alpha_1-\alpha_2$ correlation is significantly observed.

\subsubsection{Production (or measurement) rate}

Up to now, we have not distinguished in this statistical analysis between the production of an ER and its effective measurement using one type of correlation. In fact, they are linked by the appropriate branching ratio and detection efficiency. In the following, we associate $\lambda_p$ to a ''measurement rate'' which refers to an effective measurement of one type of correlation for one specific ER. 

\begin{figure*}
\includegraphics[angle=0,width=12cm]{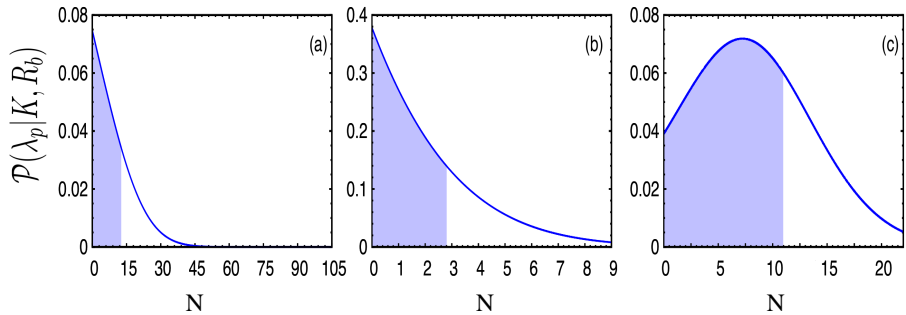}
\caption{Probability distributions for the measurement rate $\lambda_p$ of physical events for the $\{^{257}$Rf$-\alpha_1\}$ (left), $\{^{258}$Rf$-sf\}$ (middle), and $\{^{259}$Rf$-\alpha_1\}$ (right) correlations. The shaded area indicates the upper limit of $\lambda_p$ with a confidence level of $68.2\%$.}
\label{fig:lambda}
\end{figure*}

The measurement rate probability distributions are obtained from Eq.~(\ref{eq:L_runknown}) and plotted in Fig.~\ref{fig:lambda}.
An upper limit $\lambda_p^{max}(\varepsilon)$ for a measurement rate, associated to a confidence level of $100(1-\varepsilon)\%$, can be obtained from the equation
\begin{equation}
\int_0^{\lambda_p^{max}} d\lambda_p \, \mathcal{P}(\lambda_p|K,R_b)=1-\varepsilon.
\end{equation}
The shaded areas  in Fig.~\ref{fig:lambda} 
indicate the region for $\lambda_p<\lambda_p^{max}(0.318)$.
For each type of correlation, the real value of $\lambda_p$ is expected to be in this shaded area with a confidence level of $68.2\%$. The latter corresponds to the interval $\pm\sigma$ for Gaussian distributions with standard deviation $\sigma$.

For the other types of correlations where no events have been detected, the probability distribution becomes $\mathcal{P}(\lambda_p)=e^{-\lambda_p}$. In this case, the upper limit for $\lambda_p$, still with a confidence level of $68.2\%$, is $\lambda_p^{max}=1.14$ residues produced during the experiment. 
The values of $\lambda_p^{max}$ are summarized in table~\ref{tab:lambda_max}.

\begin{table}
\caption{\label{tab:lambda_max} Upper limit number $\lambda_p^{max}$ of fusion-evaporation residues produced during the experiment with a confidence level of $68.2\%$.}
\begin{tabular}{cccc}
\hline\noalign{\smallskip}
ER &  $^{257}$Rf & $^{258}$Rf & $^{259}$Rf \\
\noalign{\smallskip}\hline\noalign{\smallskip}
$\{ER-\alpha_1\}$ & 12.68 & 1.14 & 10.97 \\
$\{ER-\alpha_1-\alpha_2\}$ & 1.14 & 1.14 & 1.14 \\
$\{ER-sf\}$  & & 2.81 & \\
\noalign{\smallskip}\hline
\end{tabular}
\vspace*{0cm}  
\end{table}

\subsection{Fusion-evaporation cross-sections}

\subsubsection{Independent analysis for each type of correlation}

Using branching ratios of table~\ref{tab:ER}, together with the efficiencies and target and beam characteristics given in section~\ref{sec:exp_details}, the $\lambda_p^{max}$ of table~\ref{tab:lambda_max} are converted into upper limits for the fusion-evaporation cross-section associated to each type of correlation.  The results are shown in table~\ref{tab:csxn}. For $^{257,259}$Rf, the fact that no correlation with an emission of two alphas has been observed leads to the strongest constraint on their production cross-section.
 For $^{258}$Rf, no alpha have been observed and the correlation with only one alpha is more constraining than the one with two alphas because there is a better detection efficiency for a single alpha than for two. Note that, for this isotope, the strongest constraint is given by the $\{ER-sf\}$ correlation. 
 
\begin{table}
\caption{\label{tab:csxn} Similar to table~\ref{tab:lambda_max} for the upper limit of the cross-sections in pb.}
\begin{tabular}{cccc}
\hline\noalign{\smallskip}
ER &  $^{257}$Rf & $^{258}$Rf & $^{259}$Rf \\
\noalign{\smallskip}\hline\noalign{\smallskip}
$\{ER-\alpha_1\}$ & 554.4 & 143.1 & 502.2 \\
$\{ER-\alpha_1-\alpha_2\}$ & 172.4 & 440.8 & 235.1 \\
$\{ER-sf\}$  & & 87.2 & \\
\noalign{\smallskip}\hline
\end{tabular}
\end{table}

\subsubsection{Particular case of the $^{258}$Rf  residue}

The fact that the  $^{258}$Rf can decay by either $\alpha-$emission or spontaneous fission, with known branching ratios, can be used to slightly reduce the upper limit for the production cross-section of this isotope. 
In fact, the search for $\{^{258}$Rf $,\alpha_1\}$ and $\{^{258}$Rf $,sf\}$ correlations is equivalent to two  measurements of the $^{258}$Rf synthesis. 

Let us note $\lambda_p^{tot}$ the total production rate of this isotope, which would correspond to a measurement rate of all possible decay channels with full efficiency. Similarly, the measurement rates for this isotope with the correlations $\{^{258}$Rf$-\alpha_1\}$ and $\{^{258}$Rf$-sf\}$ (see table~\ref{tab:lambda_max}) are now noted $\lambda_p^\alpha$ and $\lambda_p^{sf}$, respectively. These quantities are related by
\begin{equation}
\lambda_p^{tot}=\frac{\lambda_p^\alpha}{\eta_\alpha}=\frac{\lambda_p^{sf}}{\eta_{sf}},
\end{equation}
where $\eta_{\alpha}$ and $\eta_{sf}$ are the product of the branching ratio and the detection efficiency for $\alpha-$decay and spontaneous-fission, respectively.
We also note $K^\alpha$ and $K^{sf}$ the total number of $\{^{258}$Rf$-\alpha_1\}$ and $\{^{258}$Rf$-sf\}$ correlations, respectively. 
Similarly, $R_b^{\alpha}$ and $R_b^{sf}$ denote the evaluation of $\{^{258}$Rf$-\alpha_1\}$ and $\{^{258}$Rf$-sf\}$ random correlations from the ''backward in time'' analysis, respectively. 

The probability density distribution of $\lambda_p^{tot}$ then reads
\begin{eqnarray}
&&\mathcal{P}(\lambda_p^{tot}|K^\alpha,R_b^\alpha,K^{sf},R_b^{sf}) \nonumber \\
&&=\frac{\mathcal{P}(\eta_\alpha\lambda_p^{tot}|K^\alpha,R_b^\alpha)\, \mathcal{P}(\eta_{sf}\lambda_p^{tot}|K^{sf},R_b^{sf})}{\int_0^\infty d\lambda\, \mathcal{P}(\eta_\alpha\lambda|K^\alpha,R_b^\alpha)\, \mathcal{P}(\eta_{sf}\lambda|K^{sf},R_b^{sf})}.
\end{eqnarray}
As a result, the upper limit for $\lambda_p^{tot}$ is 16.8 with a confidence level of $68.2\%$. 
It corresponds to a cross-section of 80.8~pb. Combining the measurements of both $\alpha-$decay and spontaneous-fission leads then to a reduction of the upper limit for the production cross-section by $\sim10\%$ as compared to the single spontaneous-fission measurement (see table~\ref{tab:csxn}).

\subsubsection{Summary}

Finally, the smallest upper limits for the production of the three rutherfordium isotopes are summarized in table~\ref{tab:cs}. Note that these final results correspond only to statistical uncertainties, i.e., no systematic errors are included.

\begin{table}
\caption{\label{tab:cs} Smallest upper limit of the cross-sections in pb, with a confidence level of $68.2\%$.}
\begin{tabular}{cccc}
\hline\noalign{\smallskip}
 residue & $^{257}$Rf & $^{258}$Rf & $^{259}$Rf \\
\noalign{\smallskip}\hline\noalign{\smallskip}
cross-section (upper limit) in pb &172.4 & 80.8 & 235.1 \\
\noalign{\smallskip}\hline
\end{tabular}
\end{table}

\section{Conclusions}
We have measured  upper limit cross-sections for the different evaporation channels of the $^{124}$Sn+$^{136}$Xe fusion reaction. 
These cross-sections are at least three orders of magnitude smaller than in the asymmetric $^{50}$Ti+$^{208}$Pb system. 
If there is no drastical (and surprising) increase of the fusion cross-section with the number of neutrons in the projectile, fusion with neutron-rich radioactive beams, even with intensity around 10$^{10}$pps, will not be sufficient to form neutron-rich transfermium and superheavy elements with the present detection setup. In addition, progress will be needed in the rejection, since the radioactive beam ions implanted in the detector could cause parasitic signals. The analysis of this experiment shows the importance of taking into account random correlation probabilities in low signal-to-noise ratio experiments, and how to handle them in a practical way. 

\section*{Acknowledgements}
We thank the Ganil beam operator who provided us a steady, intense xenon beam, and specially the technical team, G. Fr\'emont and Ch. Spitaels for their constant support during this experiment.

\end{document}